\documentclass[preprint]{aastex}
\begin{document}

\title{The Influence of Differential Irradiation and Circulation on the Thermal Evolution of Gas Giant Planets. I. Upper Limits from Radiative Equilibrium}

\author{Emily Rauscher\altaffilmark{1}}
\affil{Department of Astrophysical Sciences, Princeton University
  \\ 4 Ivy Lane, Princeton, NJ 08544, USA}
\email{rauscher@astro.princeton.edu}
\and
\author{Adam P. Showman}
\affil{Lunar and Planetary Laboratory, University of Arizona
  \\ 1629 East University Blvd., Tucson, AZ 85721, USA}
\altaffiltext{1}{NASA Sagan Fellow, Lyman P.\ Spitzer Jr.\ Fellow}

\begin{abstract}

As a planet ages it cools and its radius shrinks, at a rate set by the efficiency with which heat is transported from the interior out to space.  The bottleneck for this transport is at the boundary between the convective interior and the radiative atmosphere; the opacity there sets the global cooling rate.  Models of planetary evolution are often one-dimensional, such that the radiative-convective boundary (RCB) is defined by a single temperature, pressure, and opacity.  In reality the spatially inhomogenous stellar heating pattern and circulation in the atmosphere could deform the RCB, allowing heat from the interior to escape more efficiently through regions with lower opacity.  We present an analysis of the degree to which the RCB could be deformed and the resultant change in the evolutionary cooling rate.  In this initial work we calculate the upper limit for this effect by comparing an atmospheric structure in local radiative equilibrium to its 1D equivalent.  We find that the cooling through an uneven RCB could be enhanced over cooling through a uniform RCB by as much as 10-50\%.  We also show that the deformation of the RCB (and the enhancement of the cooling rate) increases with a greater incident stellar flux or a lower inner entropy.  Our results indicate that this mechanism could significantly change a planet's thermal evolution, causing it to cool and shrink more quickly than would otherwise be expected.  This may exacerbate the well known difficulty in explaining the very large radii observed for some hot Jupiters.

\end{abstract}

\section{Introduction}  

After a planet forms, regardless of the particular mechanism or initial conditions, it will spend the rest of its life cooling and shrinking.  The exact form of this evolution can depend on the composition of the planet \citep[as this can control the opacity of the atmosphere and the efficiency of convective heat transport;][]{Burrows2007,Leconte2012,Wood2013}, whether it is being strongly irradiated by its stellar host \citep[as this should slow its cooling;][]{Guillot1996}, and whether there are any additional processes that work to heat the interior of the planet and limit the shrinking of the radius, such as tidal dissipation \citep{Bodenheimer2001,Arras2010}, breaking of waves generated by the atmospheric circulation \citep{Showman2002}, or the ohmic dissipation of currents induced by a partially ionized atmosphere in motion \citep{Batygin2010,Perna2010b}.  These mechanisms have received much attention because it is a well known problem that many hot Jupiters have radii larger than can be explained without an additional source of heating to slow their contraction \citep[the introduction of][has a good recent review of this]{Spiegel2013}.

The cooling and evolution of a planet occurs as heat is transported from the interior of the planet out to space.  This transport begins with convective motion in the interior of the planet and then becomes radiative through the convectively stable atmosphere \citep[see][for a nice historical review of how this came to be understood for the solar system gas giants]{Hubbard1980}.  Assuming that the opacity of the atmosphere increases with pressure, the radiative transport will be the least efficient at the bottom of the atmosphere---at the radiative-convective boundary (RCB)---and this will be the bottleneck for the global cooling.  The depth of the RCB defines the cooling rate; if it is at higher pressure, the local opacity should generally be greater---and the flux through the boundary should be less---than it would be for an RCB at lower pressure.  The total cooling luminosity of the planet can be calculated by integrating over the flux through the surface of the RCB \citep{Arras2006}.

This clean picture of planet evolution is complicated by the important property that planets are not one-dimensional.  In particular, the heating planets receive from their host stars is not spatially uniform; for example, the Earth's equator is heated more than its poles.  Since hot Jupiters are expected to be tidally locked into synchronous rotation \citep[e.g.,][]{Guillot1996,Rasio1996}, there should be a huge hemispheric asymmetry between the stellar heating on their permanent day and night sides.  It has been recognized by several authors that this disparity could result in a day-night difference in the depth of the RCB (at higher/lower pressure on the day/night side) and that the more efficient cooling through the night side should increase the global cooling rate over what would be calculated through an equivalent, uniform RCB \citep{Guillot2002,Budaj2012,Spiegel2013}.\footnote{This is similar in concept to the adjustment to evolutionary models that is necessary for stars rotating quickly enough to experience gravity darkening \citep{vonZeipel1924a,vonZeipel1924b}.}  However, the same differential heating that could produce an uneven RCB also drives atmospheric circulation, which works to decrease temperature gradients and thereby influences the structure of the atmosphere.  \citet{Guillot2002}, \citet{Budaj2012}, and \citet{Spiegel2013} all recognized this effect and included parametric treatments of circulation in their 1+1D models (linking day and night side profiles).  Their results all confirmed that the greater the amount of advective day-to-night heat transport, the more uniform the RCB should be and the lower the evolutionary cooling rate.

In this work we present a new model for the deep atmospheres of hot Jupiters as a framework in which to study the deformation of the RCB through differential heating and circulation.  Our main goal of this initial study is to estimate the potential deformation of the RCB and to place quantitative upper limits on how much of a correction this effect could make to the cooling rates used in 1D evolutionary models of hot Jupiters.  We begin by describing our modeling framework, and the reasons behind our choices, in Section~\ref{sec:model}.  This includes a discussion of our radiative transfer (\ref{sec:rt}), the method we use to solve for radiative-convective equilibrium (\ref{sec:equil}), and how we calculate the global cooling luminosity (\ref{sec:lcool}).  In Section~\ref{sec:results} we present our results comparing the cooling rates from our spatially varying models to their 1D equivalents.  Our main findings are summarized in Section~\ref{sec:sum}.

\section{Deep atmosphere model} \label{sec:model}

We are unaware of any previous models that are designed to study the horizontal structure of the deep atmospheres of hot Jupiters.  While the incident stellar light should be absorbed by 1-10 bar, a hot Jupiter atmosphere is expected to remain stable against convection down to the RCB at hundreds to thousands of bars.  Standard 3D General Circulation Models (GCMs) are not well suited to model the RCB within their vertical domain for a few reasons: 1) there is a difference of about four orders of magnitude between radiative timescales at the optical (or infrared) photosphere and deep pressures \citep[$\sim$100 mbar vs.\ $\sim$100 bar, see][]{Showman2008}, making it computationally expensive to use short enough timesteps for the upper atmosphere and to integrate long enough for the deep layers to reach an equilibrium; 2) the set of simplified fluid equations solved by GCMs, the primitive equations of meteorology, assume vertical hydrostatic equilibrium and GCMs typically apply a ``convective adjustment'' to superadiabatic temperature profiles to bring them back to neutral stability; and 3) the standard bottom boundary condition for the radiative transfer is to specify a constant flux traveling up from the interior \citep[e.g.,][]{Fortney2005,Rauscher2012b}, which means that the planet's evolutionary cooling rate is defined rather than predicted.  Given the limitations of GCMs in this regard, we have chosen to develop a new modeling framework to study RCBs of hot Jupiters.

\begin{figure}[ht!]
\begin{center}
\includegraphics[width=0.5\textwidth]{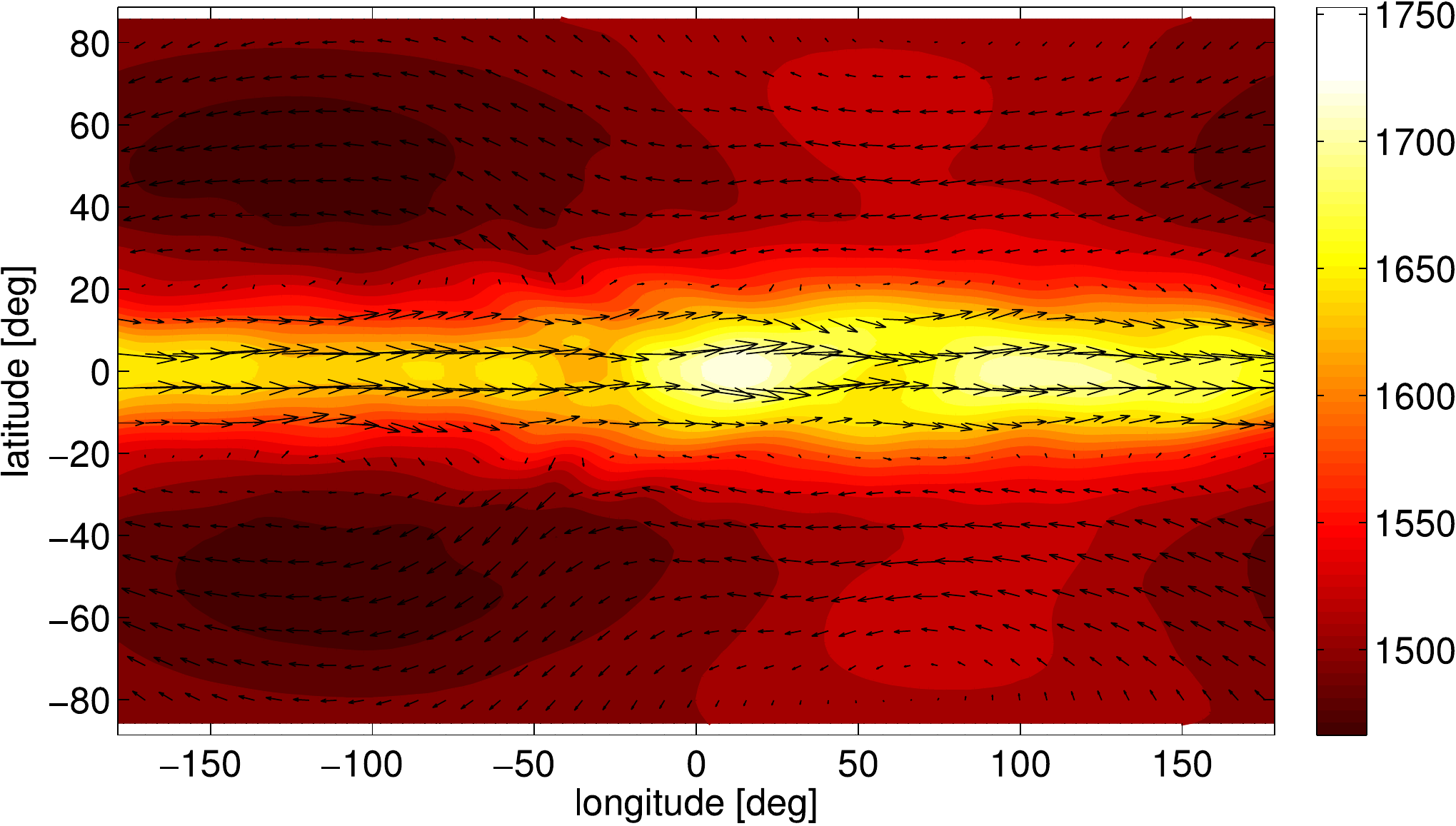}
\includegraphics[width=0.425\textwidth, trim= 0 0 0 60, clip=true]{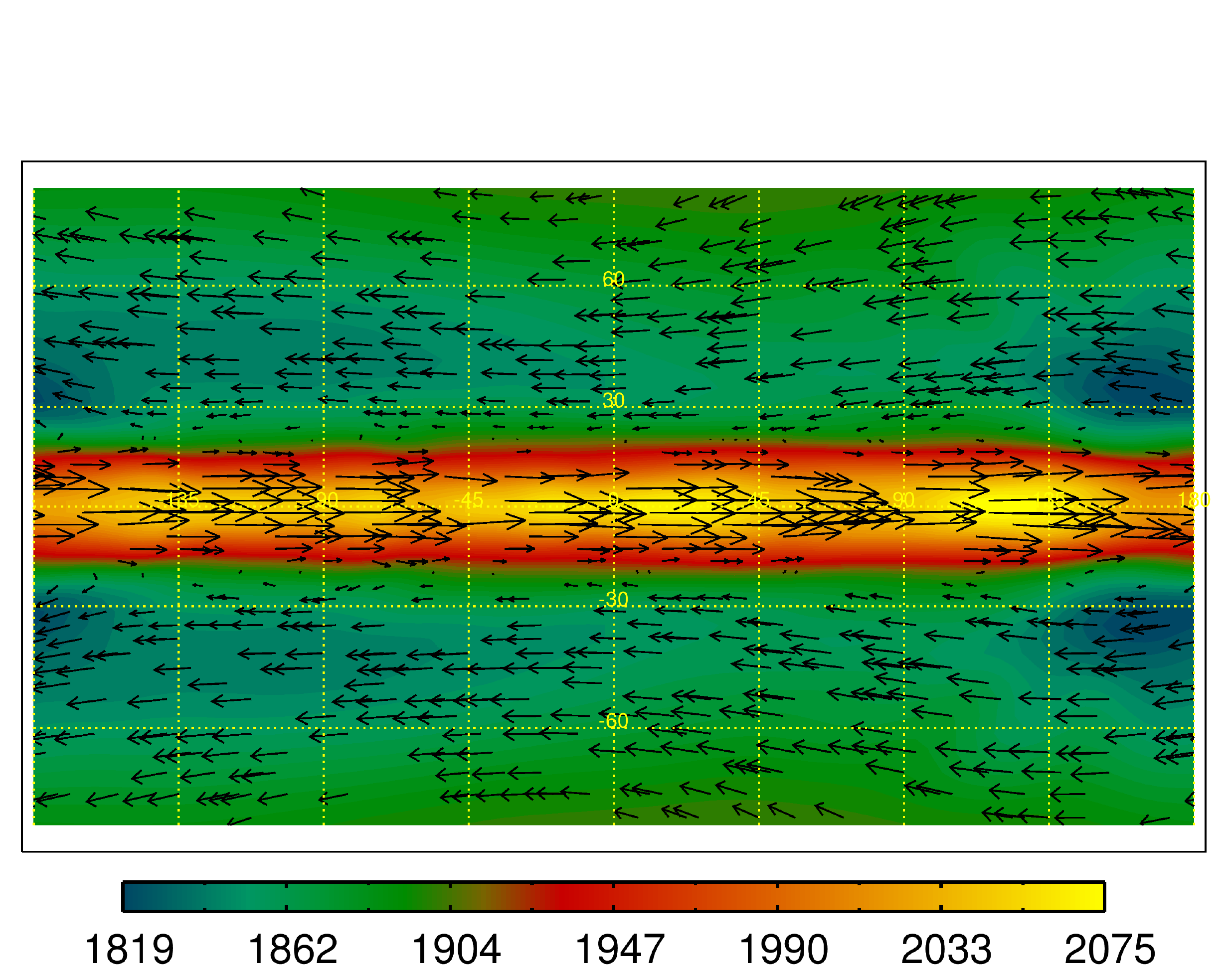}
\end{center}
\caption{Horizontal slices through the atmosphere from 3D GCMs of the planet HD 189733b (top: 10 bar level from Showman et al. 2009, bottom: 40 bar level from Rauscher \& Menou 2013), with temperature (in K) as color and winds plotted as vectors (max speeds of 1.4, 1.5 km s$^{-1}$ for the top, bottom plots).  Each map is in cylindrical projection, centered on the substellar point.  Both models assume the hot Jupiter has been tidally locked into a synchronous rotation state, such that it has permanent day and night sides.  At the deep pressures shown here, well below the optical photosphere, the temperature structure has no relation to the day-night stellar heating pattern.  It is instead characterized by a hot equator and cooler poles, a primarily axisymmetric pattern.} \label{fig:deepGCMs}
\end{figure}

We use the results of our GCMs for guidance as to the nature of the deep atmosphere and in order to choose our modeling framework.  In Figure~\ref{fig:deepGCMs} we show typical examples of hot Jupiter atmospheric structure at pressures well below the optical photosphere \citep[from][]{Showman2009,Rauscher2012b}.  At these pressures the temperature pattern is not governed directly by the strong, hemispherically uneven stellar heating, showing instead a primarily axisymmetric hot-equator/cold-pole pattern.\footnote{This structure is indirectly due to the stellar heating, however, in that the day-night forcing drives an eastward equatorial jet \citep{Showman2011} that descends throughout most of the atmosphere and shapes the temperature structure at depth.}  This can be understood in part by comparing to the shallow-water models in \citet[][see their Figures 3 and 4]{Showman2013}, which show that when the atmosphere is characterized by a long radiative time constant (as should be the case at high pressure), efficient circulation will lead to a predominantly equator-pole thermal gradient, even under constant-day/night radiative forcing conditions.  We find this same thermal structure at deep pressure levels in our 3D GCMs because the radiative transfer routines in our codes self-consistently result in long radiative timescales there.  Based on these circulation models, we choose to use a 2D, axisymmetric set-up for our deep atmosphere model.  Although the upper atmosphere is certainly not axisymmetric, here it only acts as a boundary condition for our region of interest.

In this first paper we begin by ignoring any horizontal transport of energy via atmospheric dynamics.  As discussed above, the spatial inhomogeneity of the RCB should be a balance between the differential stellar heating (acting to deform the RCB) and the response of the atmospheric circulation, evening out temperature gradients (and smoothing the RCB).  We also cited results that the less uniform the RCB, the greater the evolutionary cooling rate of the planet.  Therefore, in order to obtain an upper limit on the possible strength of this effect (taking into account the expected axisymmetry in the deep atmosphere), we consider one extreme of the radiative-dynamic balance (zero advective transport) and compare it to the opposite extreme (complete homogenization, i.e. a 1D model).  Once we understand the possible quantitative importance of uneven cooling through the RCB, we can include dynamical transport in our model and explore the role of circulation in future work.

\subsection{Coordinate system} \label{sec:coords}

We set up our axisymmetric model with latitude ($\phi$) as the horizontal coordinate and pressure ($P$) for the vertical.  In the absence of horizontal heat transport via winds,\footnote{Horizontal energy transport could occur via radiation, but winds should be dominant.  The ratio of characteristic horizontal to vertical scales in a hot Jupiter atmosphere is $\sim$10-100, so under typical conditions the horizontal divergence of the radiative flux is probably only a few percent of the vertical divergence.  Thus, even in this context, where we ignore dynamical heat transport, it is appropriate to also ignore horizontal radiative transport.} the structure of each atmospheric column can be solved for independently, by assuming local radiative equilibrium.  We solve for $N_{\mathrm{lat}}$ columns evenly spaced in latitude from the equator to the pole, with the only difference between each column being the azimuthally averaged stellar flux that heats the atmosphere.

\begin{figure}[ht!]
\begin{center}
\includegraphics[width=0.45\textwidth]{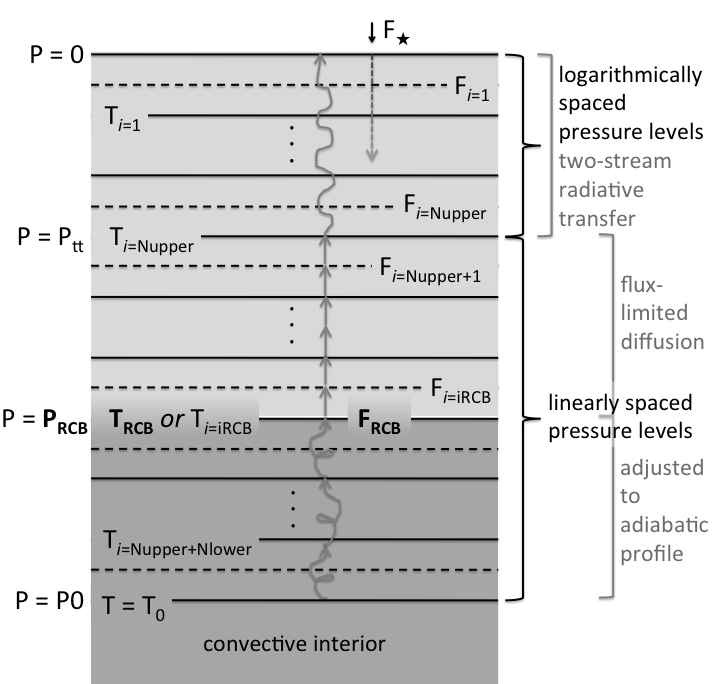}
\end{center}
\caption{A diagram of the vertical set-up for a single atmospheric column in our model.  The atmosphere is divided into $N_{\mathrm{lower}}$ linearly spaced pressure levels, the uppermost of which is at $P_{\mathrm{tt}} =P_0/(N_{\mathrm{lower}}+1)$ bar.  Above this are $N_{\mathrm{upper}}$ logarithmically spaced levels, up to 10 mbar.  We use separate radiative transfer schemes for the upper and lower portions of the atmosphere.  The fluxes are calculated at the midpoints between levels, with the stellar flux at the upper boundary ($F_{\star}$) prescribed and the flux through the RCB ($F_{\mathrm{RCB}}$) self-consistently calculated.  In the convective layers, below $P_{\mathrm{RCB}}$, no fluxes are calculated and the temperature structure is set to match the inner adiabat, defined by the bottom boundary point $[P_0,T_0]$.} \label{fig:diagram}
\end{figure}

In order to adequately resolve the upper atmosphere (near the optical and infrared photospheres), we must use logarithmically spaced pressure levels, typical of standard hot Jupiter GCMs.  However, for this application we also want to be able to resolve the location of the RCB to good precision, and so we use linearly spaced pressure levels for the deep atmosphere.  For $N_{\mathrm{lower}}$ linearly spaced pressure levels, between a bottom boundary at $P=P_0$ and an upper boundary at $P=0$, the uppermost level is at $P=P_0/(N_{\mathrm{lower}}+1)$.  We label this level $P_{\mathrm{tt}}$ and use it as the lower boundary for the $N_{\mathrm{upper}}$ logarithmically spaced pressure levels in the upper atmosphere, which are distributed evenly between $P=P_{\mathrm{tt}}$ and 10 mbar.  Figure~\ref{fig:diagram} shows a diagram of our vertical set-up.

\subsection{Radiative transfer} \label{sec:rt}

The separation of our vertical domain into upper logarithmic pressure levels and lower linear levels provides a sensible boundary at which to transition between radiative transfer schemes better suited for optically thin and thick atmospheres.  For our choice of model parameters (see Table~\ref{tab:params}), both the optical and infrared photospheres sit above $P_{\mathrm{tt}}$, which is at $\tau_{\mathrm{IR}} =74$.  The radiative transfer we use is modified from the formalism in \citet{Rauscher2012b}; more details can be found in that work, and references therein.  As is typical for many radiative transfer codes and often required for numerical stability, we calculate the flux at the midpoints between the pressure levels at which the temperatures are defined.

The strength of the stellar flux incident on the top of the planet's atmosphere is greatest at the substellar point, where it can be characterized by an irradiation temperature, $F_{\star} = \sigma T_{\mathrm{irr}}^4$, which is dependent on the star's temperature ($T_{\star}$), radius ($R_{\star}$), and distance from the planet ($d$) as: $T_{\mathrm{irr}}=T_{\star} \sqrt{R_{\star}/d}$.
Assuming a constant optical absorption coefficient ($\kappa_\star$), the stellar flux is attenuated as it travels down through the atmosphere as $F_{\star}(P) = \sigma T_{\mathrm{irr}}^4 \mu_{\star} \exp(-\kappa_{\star}P/\mu_{\star}g)$ \citep{Guillot2010}, where $\mu_{\star}$ is the cosine of the angle away from the substellar point and $F_{\star}(P)$ is only non-zero for $\mu_{\star} > 0$, \emph{i.e.} on the day side.  The $\mu_{\star}$ appears once outside the exponential to account for the decrease in flux incident on the top of the atmosphere for regions farther away from the substellar point and once within the exponential to account for the increase in path length.  Due to our assumption of axisymmetry (motivated by the arguments above), we calculate the stellar flux that travels down through each vertical column of the atmosphere as an average over all longitudes.  This disregards the day-night irradiation difference, which is fundamental to the structure of the upper atmosphere (but not our topic of study here), while preserving the necessary physics that the equator receives more stellar heating than the poles.  Thus, for a location in our atmospheric model defined by a latitude ($\phi$) and pressure ($P$), the downward stellar flux is calculated as:
\[
\bar{F}_{\star}(\phi,P) = \left. \int_{-\pi/2}^{\pi/2} F_{\star}(\phi,\lambda,P) d\lambda \middle/ \int_{0}^{2\pi} d\lambda \right.
\]
\begin{equation}
\bar{F}_{\star}(\phi,P) = \sigma T_{\mathrm{irr}}^4 \frac{\cos \phi}{2 \pi} \int_{\frac{-\pi}{2}}^{\frac{\pi}{2}} \cos \lambda 
           \exp\left( -\frac{1}{\cos \lambda \cos \phi} \frac{\kappa_{\star} P}{g} \right) d\lambda \label{eqn:vis}
\end{equation}
where we have used the overbar to emphasize that this is the flux averaged over all longitudes ($\lambda$) and have substituted $\mu_{\star}=\cos \phi \cos \lambda$.  We set the substellar point to be at $[\phi,\lambda]=[0,0]$ and only integrate the flux over the day side of the planet.

Separate from the optical flux, the infrared radiation is absorbed and re-emitted within each column according to the temperature profile.  In the upper atmosphere we use a standard two-stream gray radiative transfer scheme:\footnote{ This is the longwave broadband scheme used in the University of Reading's Intermediate General Circulation Model \citep{deForster2000}, from which the code in \citet{Rauscher2012b} is derived.  See \citet{Stephens1984}, Equation 22, for a derivation of this form for the radiative transfer.  The standard diffusivity factor of 1.66 has been used to account for the implicit integration of the isotropic thermal radiation over all upward/downward angles.}
\begin{equation}
F_{\mathrm{IR}} (P) = \int \left(1-\exp \left[ - \frac{1.66}{g} \int \kappa_{\mathrm{IR,tt}} dP \right] \right) \frac{d \sigma T^4}{dP} dP \label{eqn:thin}
\end{equation}
with a constant infrared absorption coefficient, $\kappa_{\mathrm{IR,tt}}$.  In the deep atmosphere we calculate the flux using a standard diffusion equation:
\begin{equation}
F_{\mathrm{IR}} (P) = \frac{16 g \sigma T^3}{3 \kappa_{\mathrm{IR}}} \frac{dT}{dP} \label{eqn:thick}
\end{equation}
using an infrared absorption coefficient that scales linearly with pressure and is continuous at $P = P_{\mathrm{tt}}$: 
\begin{equation}
\kappa_{\mathrm{IR}} = \kappa_{\mathrm{IR,tt}} (P/P_{\mathrm{tt}}).
\end{equation}
\citet{Guillot2010} has shown that the use of constant infrared and optical absorption coefficients can adequately model temperature-pressure profiles in the upper atmosphere, while in the lower atmosphere collision-induced absorption should introduce a linear dependence of $\kappa$ on pressure.  This is our reason for choosing to use $\kappa \propto P^0$ in the upper atmosphere and $\kappa \propto P^1$ at depth.  The choice for how the opacity scales with pressure is very important and strongly impacts our results.  For example, if we used a constant $\kappa$ in the deep atmosphere, we would find that uneven cooling through the RCB was a negligible effect, in contrast to the results we present below.  In addition, the true opacity is wavelength dependent and also a function of temperature.  The cooling through the RCB depends sensitively on the opacity there and so the reader should keep in mind that we are using a simplified form for the opacity, but we think this is an appropriate initial modeling choice given the other approximations and uncertainties present.

Mathematically, Equation~\ref{eqn:thin} transitions to Equation~\ref{eqn:thick} at high optical depth, as the exponential term trends toward zero and the flux profile becomes dependent on the local temperature gradient.  We can use Equation~\ref{eqn:thin} to check the validity of choosing $P=P_{\mathrm{tt}}$ as the boundary between our radiative transfer schemes.  Using the parameters in Table~\ref{tab:params} we calculate that the exponential term in Equation~\ref{eqn:thin}, i.e., the deviation of the flux profile from being solely dependent on the temperature gradient, to be $3\times 10^{-3}$ for the layer above $P_{\mathrm{tt}}$ and 0 for the layer below it, to the precision of our code.  Thus a transition to Equation~\ref{eqn:thick} at $P_{\mathrm{tt}}$ is quantitatively appropriate, mathematically justified, and avoids the computational error that could result otherwise \citep[see][for a discussion of this issue]{Rauscher2012b}.

We remove all layers between the RCB and the bottom boundary from the flux calculations, because the fluxes for $P>P_{\mathrm{tt}}$ are only dependent on neighboring layers and so unaffected by the deeper atmosphere.  Our aim is to predict the amount of flux emitted by the planet, in excess of the absorbed starlight.  This means that instead of defining a constant upward flux at our bottom boundary, we assume that the RCB is contained within our modeling domain, forcing the bottom boundary to necessarily be on the inner adiabat of the planet, following the precedent of 1D radiative calculations that link different regions of the planet \citep[e.g.,][]{Barman2005,Showman2008,Budaj2012}.  We fix the temperature at the boundary ($P=P_0$) to $T_0$, which is equivalent to defining the entropy of the planet interior.  The flux at the RCB can then be calculated, rather than prescribed.  By noting that the temperature gradient at the RCB is necessarily adiabatic, the flux there is easily calculated as \citep{Arras2006}:
\begin{equation}
F_{\mathrm{RCB}} = - \frac{16 \sigma g (\mathcal{R}/c_p) T_{\mathrm{RCB}}^4}{3 \kappa_{\mathrm{IR,RCB}} P_{\mathrm{RCB}} }.
\end{equation}

\subsection{Solving for radiative-convective equilibrium} \label{sec:equil}

We have described the manner in which we solve for fluxes throughout the atmosphere, given a temperature structure, but in order to find a steady-state solution we must invert the problem to solve for the temperature structure that gives a constant upward flux ($dF/dP=0$), equal to the flux at the RCB, in each column of the atmosphere.  To do this, we employ a method similar to the one described in \citet{McKay1989}.  From an initial guess for the temperature profile in each column, we calculate the net fluxes at the midpoints between each temperature level, $F_i$, using the schemes described above.  We find the difference between these fluxes and the flux through the RCB for that column (the initial profile is convective in its deepest levels), giving us a vector of residuals, $R_i = F_i - F_{\mathrm{RCB}}$.  We perturb the temperature at each level and recalculate the fluxes in order to solve for each term in the array: $A_{ij} = dF_i/dT_j$.  A standard matrix solver gives the adjustments ($\delta T_i$) to the temperature profile such that $-\vec{R} = A \times \delta \vec{T}$, meaning that in the adjusted profile all net fluxes should be equal to $F_{\mathrm{RCB}}$.

All of the convective layers below the RCB are excluded from the flux calculations and array inversion, but the temperature at the RCB does affect the flux in the layer above it and so is included.  If the matrix solver calculates a positive adjustment to the temperature at the current RCB level, $\delta T_{i=iRCB} > 0$, the adjustment is applied and the location of the RCB is shifted down by one level, $iRCB = iRCB +1$.  If the matrix solver calculates a negative adjustment for $T_{i=iRCB}$, this would bring the local temperature below the inner adiabat and so the adjustment is ignored.  However, if the layers above the current RCB are cooled to the point that the lapse rate matches or exceeds the adiabatic lapse rate ($d\ln T/d\ln P \ge \mathcal{R}/c_p$), then those layers are adjusted to match the inner adiabat ($T_i = T_0 (P_i/P_0)^{\mathcal{R}/c_p}$) and the RCB is shifted up appropriately ($iRCB$ is decreased by however many contiguous layers were adjusted).  We then repeat the calculation of fluxes and solve for the next set of temperature adjustments using the new $F_{\mathrm{RCB}}=F_{i=iRCB}$.  We continue to iterate in this manner until the code has settled on the RCB level that gives the lowest value for the root-mean-square of the residuals, with the added requirement that this value is also less in magnitude than $F_{\mathrm{RCB}}$.

\subsection{Calculating the global cooling luminosity} \label{sec:lcool}

Once we have our solution for the 2D temperature structure, we integrate over all latitudes to calculate the cooling luminosity predicted by our model:
\begin{equation}
L_{\mathrm{cool,2D}} = 4 \pi R_p^2 \int_0^{\frac{\pi}{2}} \bar{F}_{\mathrm{RCB}}(\phi) \cos \phi \ d\phi . \label{eqn:L2D}
\end{equation}
(Since we assume north-south symmetry and only solve for one hemisphere, we integrate from the equator to pole and multiply by two.)  We compare this to the cooling luminosity predicted by a 1D model that we calculate using our same code, but for a single column with a globally averaged stellar flux function: $\bar{F}_{\star}(P) = 0.25 \sigma T_{\mathrm{irr}}^4 \exp ( -\sqrt{3} \kappa_{\star} P/g)$ \citep{Guillot2010}.  For a uniform RCB, defined by a single temperature and pressure, the cooling luminosity is simply \citep{Arras2006}:
\begin{equation}
L_{\mathrm{cool,1D}} = (64 \pi R_p^2 g/3) (\mathcal{R}/c_p) (\sigma T^4/\kappa_{\mathrm{IR}} P)_{\mathrm{RCB}}.
\end{equation}
In this equation, as in Equation~\ref{eqn:L2D} above, we assume that $R_{\mathrm{RCB}}=R_p$.  Since our results are given as $L_{\mathrm{cool,2D}}/L_{\mathrm{cool,1D}}$ this assumption is irrelevant because the radii cancel out.  However, we have neglected the variation with latitude of $R_{\mathrm{RCB}}$ in Equation~\ref{eqn:L2D}.  The effect of this approximation is to artificially decrease $L_{\mathrm{cool,2D}}/L_{\mathrm{cool,1D}}$; for typical hot Jupiter parameters we estimate this to be an effect at the $\sim$1\% level.

We have performed resolution tests on our code, in pressure and latitude, in order to determine how many grid points are necessary to accurately calculate the global cooling luminosity.  There is no clear trend in $L_{\mathrm{cool,2D}}/L_{\mathrm{cool,1D}}$ with increasing resolution; instead we see a variation at the level of $\sim$5\%.  We therefore take this to be our expected computational error and use a resolution of: $N_{\mathrm{upper}}=60$, $N_{\mathrm{lower}}=299$, and $N_{\mathrm{lat}}=20$.

We also test the temperature-pressure profiles that we find against the analytic profiles from \cite{Guillot2010}, modified for a non-constant IR absorption coefficient \citep[see Appendix A of][]{Rauscher2012b}.  However, the comparison is not exact for several reasons: 1) our profiles transition between a constant $\kappa_{\mathrm{IR}}$ and one that varies linearly with pressure, 2) the Guillot profiles do not include the effect of convection and so become superadiabatic at depth, 3) the profiles use a prescribed upward flux at the lower boundary, and 4) the profiles are calculated as a function of the angle away from the substellar point, rather than as an average over longitude.  Nevertheless, a comparison with equivalent profiles for the night side and the substellar point demonstrate that our profiles stay within these limits, and a comparison with the analytic global average demonstrates that we have successfully reproduced the general shape and magnitude of the profile with our code.

\section{Results} \label{sec:results}

\begin{figure}[ht!]
\begin{center}
\includegraphics[width=0.47\textwidth]{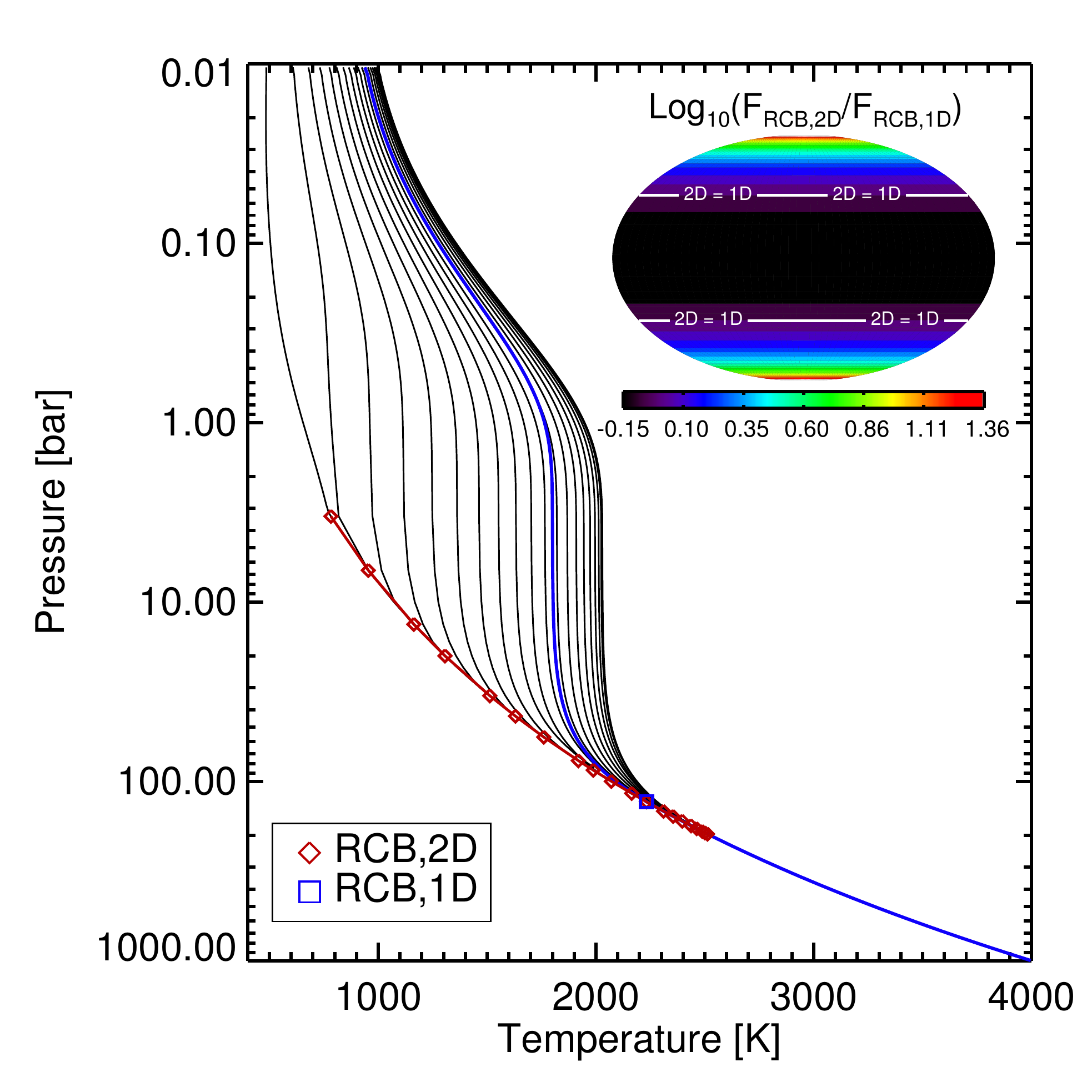}
\end{center}
\caption{A representative local-radiative-equilibrium structure calculated by our code, for $T_0=4000$ K and $T_{\mathrm{irr}} = 1500$ K.  We plot the temperature-pressure profile for each atmospheric column, from near the equator (hottest profile) to near the pole (coldest profile), with a red diamond marking the RCB for each profile.  The blue line shows the 1D, globally averaged profile we calculate, with its RCB marked by a blue square.  Inset is a map (in an area-conserving projection) showing the flux through the RCB for our axisymmetric model, normalized to the flux through the RCB for a 1D model (and using a logarithmic scale).  Although the flux out through the poles is enhanced by over an order of magnitude, the flux near the equator is slightly suppressed and this area covers a large fraction of the surface.  The integrated cooling luminosity from this 2D model is 43\% greater than its 1D equivalent.} \label{fig:tpprofile}
\end{figure}

We choose physical parameters to represent a generic hot Jupiter, such that our temperature-pressure profiles are within the range calculated for specific planets from 1D radiative-convective models with complex radiation transfer \citep[e.g.,][]{Barman2005,Fortney2005,Iro2005,Burrows2006}, following the approach of \citet{Guillot2010}.  These values are listed in Table~\ref{tab:params} and a representative set of temperature-pressure profiles are shown in Figure~\ref{fig:tpprofile}.  For reasonable parameter choices, slightly varying the chosen opacities or gravitational acceleration will shift the isothermal portion of a profile in pressure and change its extent.  An increase of the optical absorption coefficient, relative to the infrared coefficient, would produce a temperature inversion in the upper atmosphere \citep{Guillot2010}, but the presence of a day-side temperature inversion should not strongly modify the global evolution of a planet with day-night heat transport \citep{Spiegel2013} and so we do not consider this case.

\begin{deluxetable}{lcc}
\tablewidth{0pt}
\tablecaption{Model parameters}
\tablehead{
\colhead{Parameter}  &  \colhead{value} &  \colhead{units}
}
\startdata
Ratio of gas const.\ to heat capacity, $\mathcal{R}/c_p$  & 0.286  & -  \\ 
Gravitational acceleration, $g$        &   15   &   m~s$^{-2}$ \\ 
Pressure at upper boundary &  0.01  &  bar \\ 
Pressure at upper/lower atm.\ boundary, $P_{\mathrm{tt}}$  &  3.33   &  bar \\
Pressure at bottom boundary, $P_0$  &  1000 & bar \\
Optical absorption coefficient, $\kappa_\star$   &   0.002   &   cm$^2$~g$^{-1}$  \\
Optical photosphere ($\tau_\star= 2/3$)  & 0.5 &  bar \\ 
IR absorption coefficient at $P_{\mathrm{tt}}$, $\kappa_{\mathrm{IR,tt}}$  & 0.033 & cm$^2$~g$^{-1}$ \\
IR photosphere ($\tau_{\mathrm{IR}} = 2/3$)  & 320 & mbar
\enddata
\label{tab:params}
\end{deluxetable}

In this initial paper we focus on quantifying the maximum possible importance of an uneven RCB on a planet's global evolution, within this axisymmetric framework.  The most important parameter to study is therefore $T_0$, the temperature at the bottom boundary, which is directly related to to the entropy of the planet interior.  As a planet cools and contracts, $T_0$ will decrease and the RCB will move to deeper pressures, where the opacity is higher and therefore the global cooling rate will be slower.  Within the context of our 2D model, a deeper RCB should be more prone to distortion, as is demonstrated by considering the limiting extremes.  In the limit of very high entropy in the interior, where the flux out through the planet is much greater than the incident stellar flux, the RCB should be at very low pressure and there should be no significant difference between the location of the RCB at the equator and the poles.  In the other limit, where the stellar irradiation dominates over the flux from the interior, the RCB should be at very high pressures and we expect a maximal deformation of the RCB between the equator and poles.  This means that we expect the effect of uneven cooling through the RCB to matter more for lower interior entropy or, equivalently, later in a planet's lifetime.

\begin{figure}[ht!]
\begin{center}
\includegraphics[width=0.45\textwidth]{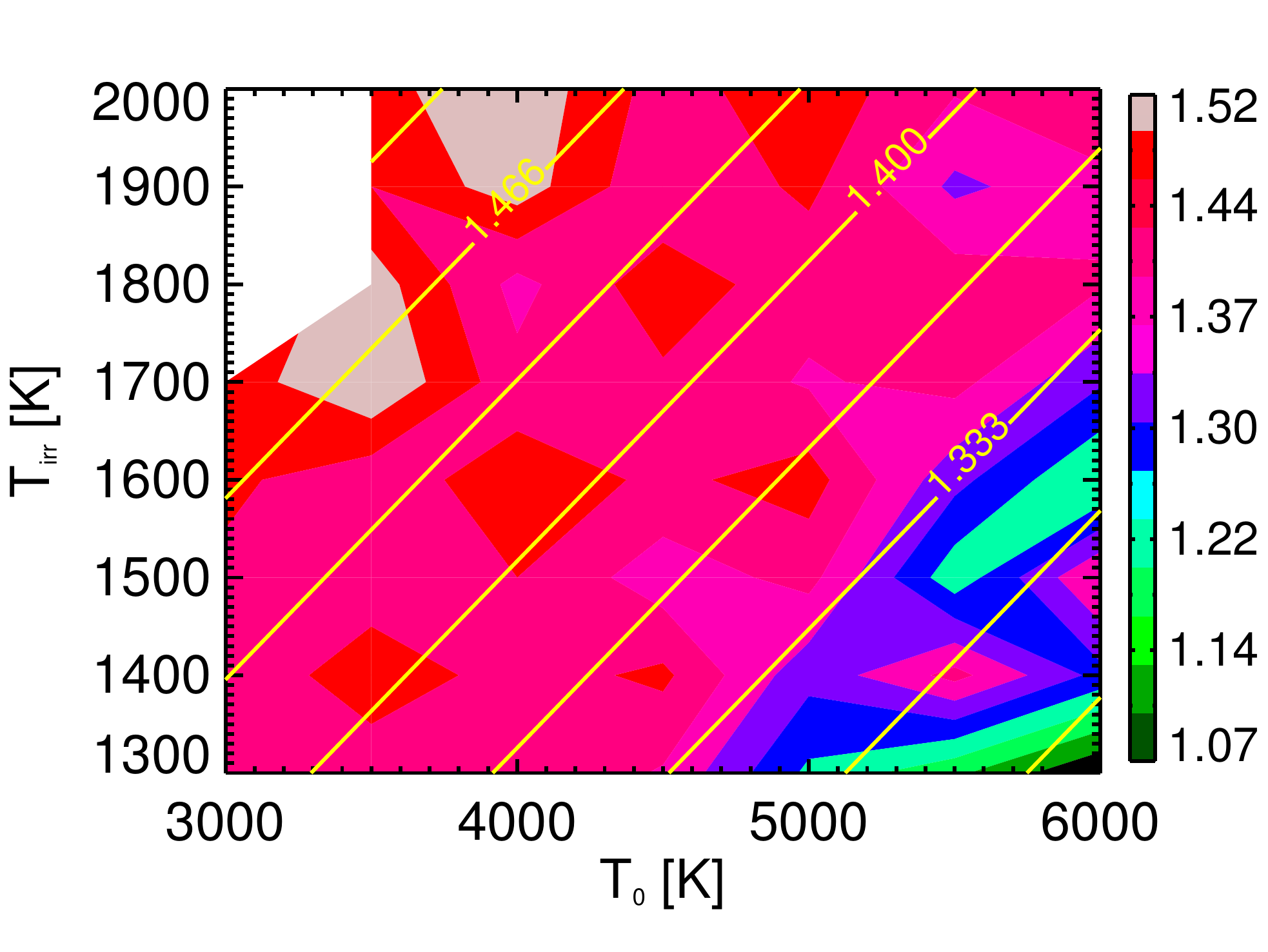}
\end{center}
\caption{The 2D correction to 1D model cooling rates, $L_{\mathrm{cool,2D}}/L_{\mathrm{cool,1D}}$, for radiative-convective-equilibrium models, as a function of internal entropy (characterized by $T_0$) and stellar irradiation (characterized by $T_{\mathrm{irr}}$).  (Models in the upper left corner had RCBs below our bottom boundary.)  Numerical noise at the level of $\sim$5\% can be seen as local variations in this parameter space.  Above the noise is a general trend: the cooling through an uneven RCB becomes increasingly important for lower internal entropy ($T_0$) and higher stellar heating ($T_{\mathrm{irr}}$).  This trend is roughly fit as a linear function of $T_0$ and $T_{\mathrm{irr}}$ (yellow contours).} \label{fig:coolingrates}
\end{figure}

This is in fact the trend confirmed by our results, shown in Figure~\ref{fig:coolingrates}.  While a young, hot planet may have a fairly uniform RCB, such that its 2D global cooling rate is similar to its equivalent 1D value, an older and cooler planet can have a strongly distorted RCB.  The 2D global cooling rate may be increased over the value from a 1D model by as much as 50\%.

The other result we see in Figure~\ref{fig:coolingrates} is that the distortion of the RCB is greater for planets that are subject to more intense stellar heating.  Just as a cooler $T_0$ pulls the RCB to deeper pressure levels, so too does a greater stellar flux, $F_{\star}=\sigma T_{\mathrm{irr}}^4$.  Although we cannot measure the current inner entropy of observed hot Jupiters, we do know the amount of incident flux they receive from their host stars.  For example, all else being equal, we would expect the enhancement of cooling through an uneven RCB to matter more for the bright hot Jupiter HD~209458b ($T_{\mathrm{irr}} \sim2100$ K) than the equally well-known HD~189733b ($T_{\mathrm{irr}} \sim1700$ K).

Our results confirm that a non-uniform RCB should enhance the cooling rate of a planet and speed its evolution \citep{Guillot2002,Budaj2012,Spiegel2013}.  This exacerbates the issue, mentioned above, of hot Jupiters with unusually large radii.  If a planet cools more quickly than would be expected from a 1D model, then whatever source of additional heating acts to keep the planet from contracting must be even stronger than would be predicted by the 1D model.  Our results continue to further complicate this issue.  Observations indicate that the more inflated planets are the ones subject to more intense stellar irradiation \citep{Demory2011,Laughlin2011,Miller2011}, but we have shown that those are the planets that could have more strongly deformed RCBs and therefore more enhanced global cooling rates, working in opposition to the observed trend.

In order to help quantify our results, we fit the values plotted in Figure~\ref{fig:coolingrates} with a simple function, linear in $T_0$ and $T_{\mathrm{irr}}$:
\begin{equation}
\frac{L_{\mathrm{cool,2D}}}{L_{\mathrm{cool,1D}}} = 1.35 
+ 0.23 \left(\frac{T_{\mathrm{irr}}}{1300\ \mathrm{K}}\right) - 0.16 \left(\frac{T_0}{3000\ \mathrm{K}}\right).
\end{equation}
This should not be viewed as a detailed prediction (it is certainly not reliable to better than $\sim$10\%), but instead a way of roughly quantifying the strength of uneven cooling and its dependence on two important parameters.  In particular, the exact values of the constant and coefficients are likely to depend on particular details of any specific planet (such as its composition, opacities, and gravitational acceleration).

There are important caveats and subtleties related to this work that should be noted.  For one, in Figure~\ref{fig:coolingrates} we are comparing models that all have the same gravitational acceleration, $g$.  This means that we cannot exactly map a line of constant $T_{\mathrm{irr}}$ to the evolutionary path of a planet (as $T_0$ decreases with time), since the radius should be decreasing and $g$ increasing.  Tests comparing models with constant $T_{\mathrm{irr}}$ and $T_0$ show that a variation of $g$ by a factor of two can change the value of $L_{\mathrm{cool,2D}}/L_{\mathrm{cool,1D}}$ by a few to 10\%.  This is about equal to, or less than, the strength of other effects that should matter; in particular, it may be smaller than the expected variation of opacity with temperature \citep{Arras2006}.

In addition, we have assumed that the convective interior of the planet is homogeneous and can be described by a single adiabat, ignoring any complexity in the temperature profile or heat transport such as that involved in double-diffusive convection \citep{Leconte2012,Wood2013} or horizontal transport of energy within the convective interior \citep{Ingersoll1978}.  We also neglect any deformation of the RCB through convective overshoot or other 3D, dynamic processes.  Given our initial modeling framework, these complexities are justifiably left for future work.

\section{Summary} \label{sec:sum}

We have performed an initial study of the deep atmospheres of hot Jupiters, modeling spatial variation of the radiative-convective boundary (RCB).  Instead of the RCB being defined by a single temperature, pressure, and opacity, the differential heating of the atmosphere can deform its surface.  This inhomogeneity leads to a global cooling rate increased from what would be calculated (as in a 1D model) for a single-valued RCB, speeding the evolution of a planet and the shrinking of its radius.  The main point of this paper is to quantify the maximum possible deviation of the cooling rate through an uneven RCB from the value that would be expected for a uniform RCB, within our 2D axisymmetric modeling framework.  In other words, we are calculating the maximum correction that might need to be applied to a 1D evolutionary model that neglects this effect.  

Based on our previous results from 3D atmospheric circulation models, we expect that the deep atmosphere is predominantly axisymmetric, such that a 2D, azimuthally averaged temperature profile should be a good representation of the full 3D structure.  We use this framework to calculate the upper limit on the uneven RCB cooling rate by considering the extreme situation in which atmospheric winds do not decrease temperature gradients on the planet and the atmospheric structure is set by assuming local radiative equilibrium.  Although this initial calculation has neglected some of the physical complexity that should be relevant to this problem (e.g., the detailed dependence of opacity on temperature and pressure), we have three robust results:
\begin{itemize}
\item We confirm that the presence of an uneven RCB increases a planet's cooling flux relative to an equivalent 1D model with the same internal entropy and incident stellar flux ($L_{\mathrm{cool,2D}}/L_{\mathrm{cool,1D}} > 1$).
\item We calculate that the uneven cooling rate could be as much as 10-50\% greater than the equivalent cooling rate through a uniform RCB.
\item We show that the enhancement of the cooling rate increases with greater incident stellar flux (characterized here by $T_{\mathrm{irr}}$) and decreases for planets with greater internal entropy (characterized here by $T_0$).
\end{itemize}

Our findings indicate that this effect could be strong enough to significantly alter the evolutionary cooling and contraction of planets, as calculated by 1D models, and therefore deserves further attention.  The next important step will be to include the influence of atmospheric circulation on the spatial variation of the RCB.  We also recognize the additional importance of working on this issue because it may exacerbate the well known problem of hot Jupiters with radii larger than predicted by 1D evolutionary models, unless there is an additional source of heating in the interior of the planet.

\acknowledgements

We thank David Spiegel for helpful input and the anonymous referee for comments that improved the quality of this paper.   This work was performed in part under contract with the California Institute of Technology (Caltech) funded by NASA through the Sagan Fellowship Program.  APS was supported by NASA Origins grant NNX12AI79G.

\bibliography{biblio.bib}  

\end{document}